\documentclass[conference]{IEEEtran}
\IEEEoverridecommandlockouts
\usepackage{cite}
\usepackage{amsmath,amssymb,amsfonts}
\usepackage{algorithmic}
\usepackage{graphicx}
\usepackage{textcomp}
\usepackage{xcolor}
\usepackage{multirow}
\usepackage{enumitem}
\usepackage{booktabs}
\usepackage{colortbl}
\usepackage{color,soul}
\usepackage{subcaption}
\usepackage[hyphens]{url}

\usepackage[ruled,vlined]{algorithm2e}

\def\BibTeX{{\rm B\kern-.05em{\sc i\kern-.025em b}\kern-.08em
    T\kern-.1667em\lower.7ex\hbox{E}\kern-.125emX}}
\begin{document}


\title{Jointly Optimizing Sensing Pipelines for Multimodal Mixed Reality Interaction}

\author{
\IEEEauthorblockN{
Darshana Rathnayake\IEEEauthorrefmark{1},
Ashen de Silva\IEEEauthorrefmark{2},
Dasun Puwakdandawa\IEEEauthorrefmark{2},
Lakmal Meegahapola\IEEEauthorrefmark{3}\IEEEauthorrefmark{4},
Archan Misra\IEEEauthorrefmark{1},
Indika Perera\IEEEauthorrefmark{2}
}
\IEEEauthorblockA{
\IEEEauthorrefmark{1}Singapore Management University, Singapore\\
\IEEEauthorrefmark{2}University of Moratuwa, Sri Lanka\\
\IEEEauthorrefmark{3}Idiap Research Institute, Switzerland\\
\IEEEauthorrefmark{4}École Polytechnique Fédérale de Lausanne (EPFL), Switzerland\\
}}

\maketitle


\maketitle

\begin{abstract}
    Natural human interactions for Mixed Reality Applications are overwhelmingly multimodal: humans communicate intent and instructions via a combination of visual, aural and gestural cues. However, supporting low-latency and accurate comprehension of such multimodal instructions (MMI), on resource-constrained wearable devices, remains an open challenge, especially as the state-of-the-art comprehension techniques for each individual modality increasingly utilize complex Deep Neural Network models. We demonstrate the possibility of overcoming the core limitation of latency--vs.--accuracy tradeoff by exploiting cross-modal dependencies--i.e., by compensating for the inferior performance of one model with an increased accuracy of more complex model of a different modality. We present a sensor fusion architecture that performs MMI comprehension in a quasi-synchronous fashion, by fusing visual, speech and gestural input. The architecture is reconfigurable and supports dynamic modification of the complexity of the data processing pipeline for each individual modality in response to contextual changes. Using a representative ``classroom” context and a set of four common interaction primitives, we then demonstrate how the choices between low and high complexity models for each individual modality are coupled. In particular, we show that (a) a judicious combination of low and high complexity models across modalities can offer a dramatic 3-fold decrease in comprehension latency together with an increase $\sim$10-15\% in accuracy, and (b) the right collective choice of models is \emph{context dependent}, with the performance of some model combinations being significantly more sensitive to changes in scene context or choice of interaction.
\end{abstract}


\section{Introduction}
 The rapid growth in the sensing capabilities of mobile and wearable devices, together with advances in machine learning-based perception, has spawned growing interest in the \emph{Mixed Reality (MR)} applications across various domains~\cite{LAMPROPOULOS202032, Meegahapola2017}. Broadly speaking, MR seeks to present users with a richer interaction and instructioning capability over a combination of both (a) synthetically-generated virtual objects, and (b) real-world objects located in the user's physical world. MR-based interaction now encompasses both unimodal~\cite{Saroha2011} and multimodal~\cite{Jamali2014} paradigms.  In general, unimodal interaction (e.g., via touch-screen interactions) is easier to parse, but is less natural, for two reasons: (a) human instructioning and interaction is inherently multimodal, employing voice, vision, gestures and touch, and (b) the constraints of unimodal technologies often imply the use of unnaturally longer sequence of actions \& instructions. Indeed,  certain modalities may be better adapted for certain types of interaction tasks--for example, speech is better for describing object attributes such as type and color, whereas gestures are more powerful for disambiguating object location.  
 
To enable truly interactive MR applications, it is important to support such multimodal instructioning and interaction capability in a real-time fashion--e.g., with a system response latency that does not exceed 1-2 secs. \emph{Multi-modality} has thus emerged as a powerful paradigm for improving the effectiveness of real-time MR interaction, with a large body of work demonstrating that multi-modality can (a) enhance the overall expressiveness of interactions~\cite{Lee2013}, (b) reduce overall interaction time~\cite{Lee2013}, (c) increase interaction accuracy~\cite{Nizam2018} and (d) support more natural interaction~\cite{Nizam2018}. Moreover, multi-modal interactions (\textbf{MMI}) are very effective in overcoming the natural ambiguity in intent expression that exists in unimodal systems. For example, in a study room scenario where several people are in the same MR realm, instructional ambiguity can arise due to imprecise perspectives~\cite{scalise}, such as the speech command ``look at \textit{this} book'' which involves the deictic expression \textit{this}. Using further verbal elaboration of ``\textit{this}'' to overcome this ambiguity is decidedly unnatural and requires more human effort.  An intuitive and alternative disambiguation approach  is to introduce another modality (such as a pointing gesture directed towards the object referred to by the ``this'' qualifier)~\cite{williams2018framework}.  This can also be viewed as providing more contextual evidence, with the gestural modality effectively providing additional context~\cite{ubiqtalker} in addition to linguistic cues. 

Most unimodal sensing and comprehension pipelines, however, explicit a natural \emph{latency vs. accuracy} tradeoff, that poses a challenge to our natural desire for low-latency, accurate interaction. This tradeoff problem has exacerbated with the recent explosion of Deep Neural Network (DNN) models for perception tasks in vision and speech ~\cite{Redmon2016, Hannun2014}--while such DNNs can increase accuracy significantly, they are often impossible to execute on resource-constrained wearable devices and must be offloaded to a GPU-rich, cloud infrastructure which imposes non-trivial additional network latency. 


Our work in this paper explores the implications of such performance tradeoffs in MMI-based MR scenarios, with a view to developing techniques to ``flatten this accuracy-vs-latency curve"--- effectively, allowing wearable devices to exploit the enhanced accuracy of multimodality without suffering the penalty of significantly longer execution latency. More specifically, the work is driven by two observations:
\begin{itemize}[leftmargin=*]
\item Current MMI designs perform algorithm selection of individual modalities in isolation, rather than jointly. In other words, the choice between a (more complex, higher-delay) DNN model vs. a (less complex, lower latency) alternative for visual object recognition is determined independently of similar choices made for other modalities. This is arguably sub-optimal as it ignores the possible interactions between different modalities, and the possibility that errors in one particular modality might be sufficiently compensated by improved capabilities of another modality.
\item Besides not being optimized jointly, the current algorithm choices are also \emph{not adaptive}--i.e., they typically tend to have a predefined fusion logic across different input modalities, independent of \emph{context}. For example, for an MR application that combines audio and gestural cues, it will continue to execute a pre-defined fusion process even if, in certain situations (e.g., a very sparse, spatially well-separated layout of books in the aforementioned study room), the gestural input may have sufficient discriminative ability, making high verbal comprehension accuracy unnecessary. 
\end{itemize}

Our contribution is to propose a sensor fusion architecture for such interactive MR applications, targeted to resource-constrained wearable devices, that can address the above-mentioned limitations. We design an architecture that is \emph{configurable}--i.e., it allows the different sense-making pipelines for each individual modality to be configured or modified, to better match (a) different types of environmental context (e.g. classrooms, conference rooms, and industrial settings), and (b) varying performance characteristics of the underlying system (e.g. clock speed, RAM size and network latency). The architecture is designed to handle the \emph{asynchronous} execution of each modality's perception pipeline. While our proposed architecture is generic, we specifically instantiate and evaluate it for fusing three common modalities: \emph{vision, speech, and gestures}. Using state-of-the-art DNN-based pipelines for each modality, we experimentally show that the judicious \emph{joint selection} of modality-specific perception pipelines  helps to significantly improve the MMI latency-vs-accuracy tradeoff.

\noindent \textbf{Key Contributions:} This paper makes the following key contributions:
\begin{itemize}[leftmargin=*]
\item \emph{Configurable Multi-modal Fusion Architecture:} We propose an asynchronous sensor fusion approach for multimodal instruction comprehension (a key building block of MMI), which combines the inferences from different modalities in a flexible fashion, while allowing the sense-making pipelines of individual modalities to be adaptively modified in response to changing environmental or device context.  The proposed architecture utilizes soft-synchronization via communication queues to coordinate inputs across different modalities, while accommodating the differences in processing latency for each modality-specific inferencing pipeline. 

\item \emph{Demonstrate Coupled MMI Tradeoffs across Modalities:} By profiling a multiplicity of different inferencing models (varying from low to high complexity), we first characterize the accuracy vs. latency characteristics for each modality.  We subsequently perform multimodal fusion, under these varying low$\leftarrow \rightarrow$high complexity alternatives, to quantify the overall impact on the performance of multimodal instruction comprehension. We show that the intelligent exploitation of compensatory synergies across different modes has significant impact: for example, (a) we can reduce latency by 10+\% without any loss of accuracy, and (b) maintain high comprehension accuracy while lowering the CPU utilization by $\sim$7\%.

\end{itemize}
We believe that our work provides tangible evidence of the benefits of a adaptive  fusion model for pervasive multimodal instruction comprehension, and should motivate future work on developing \emph{context-adaptive} MMI systems that can improve the efficiency of instruction comprehension by dynamically adapting the individual sense-making pipelines.

\section{Related Work}
We describe past work on both (a) mixed reality interactions and (b) the broader topic of multimodal sensor fusion.

\subsection{Mixed Reality Interactions}

\noindent \textit{Auditory Modality:} Hughes et al. \cite{mr-in-environment} emphasized the use of audio/speech as a modality to interact with non-linear MR environments, highlighting the importance of employing 3D surround technologies to enrich immersive experiences. Audio/speech interactions are vital to natural MR interactions, as it has been demonstrated~\cite{cavazza2004multimodal} to be the primary modality for natural human communication. Use of different auditory-related technologies such as surround \cite{Duraiswami2004}, binaural \cite{Yao-binaural-headphones} and 3D \cite{mr-in-environment}, enrich the user experience of MR applications. 


\noindent \textit{Gesture Modality:} Gestures have been widely used to convey (or clarify) human intent, in a variety of MR applications. In particular, gestures such as pointing, grabbing, or stretching have been shown to increase the immersiveness of these systems~\cite{lee2010hand}. Contact- and vision-based devices are the two main technologies in gesture recognition systems~\cite{Kareem2012}. A physical interfacing device captures the interaction of the user in the systems that have employed a contact-based device. The latter approach captures gestures with different kinds of cameras, e.g., depth camera, stereo camera, web-camera, etc.




\noindent \textit{Visual Modality:} In MR realms, visual modality is especially important as it displays both digitally synthesized and real-world physical objects, permitting seamless interaction with both object types~\cite{mr_mode}. A key factor for an effective MR experience is modeling a seamless boundary between the real and virtual objects, so that interacting with either presents no discernible difference \cite{magicbook}. Technologies such as see-through screens and opaque HMDs make the visual interactions possible in various application domains, e.g., education \cite{liarokapis2002multimedia} and training \& simulation \cite{training}.


\subsection{Multimodal Sensor Fusion}
There is a significantly large body of academic work in the past on the fusion of different sensing modalities to extract a variety of insights in different domains \cite{Jayarajah2018}. In~\cite{123}, Sargin et al. applied Hidden Markov Models (HMM) to interpret the correlation between gestures (for example, head gestures) and speech. Asano et al. presented a Bayesian network-based approach for detecting speech events, using both 1) sounds captured by an array of microphones and 2) human gestures captured by video sensors~\cite{1257294}. A similar use case was investigated in \cite{4415116}, to perform audio localization and thereby separate multiple speakers from one other and from surrounding noise. Hara et al.~\cite{123123} also present a similar solution, using Bayesian Networks to isolate and extract speech utterances by specific individuals involved in interactions with robotic agents. Hol et. al.~\cite{122122} apply Kalman filters to fuse trajectories estimated from inertial and vision sensors, to accurately identify camera dynamics (such as orientation and position) for AR applications. These techniques are leveraged in applications presented in \cite{1087440}, which primarily fuse visual and auditory cues for a natural human-agent interaction. 


\section{Proposed Approach for MMI-driven Sensor Fusion}

Sensor fusion methodologies can be categorized into multiple types~ \cite{Sharma1998}: (a) data fusion combines data from several sources; (b) feature fusion extracts features from various modalities and embeds them into a composite feature map for final inference; while (c) decision fusion involves fusing decisions generated by multiple independent inferencing models. To support the fusion of visual, speech, and gestural modalities (the 3 input modes most commonly used in natural interactions), we propose an architecture that consists of multiple independent subsystems (engines). The Vision, Aural, and Gestural engines perform modality-specific inferencing tasks, including object detection, speech recognition and text classification, and gesture recognition, respectively. The final outcome of a user interaction is based on a \emph{decision} fusion of the output of these individual inferencing pipelines.

\begin{figure}
  \centering 
  \includegraphics[width=0.95\linewidth]{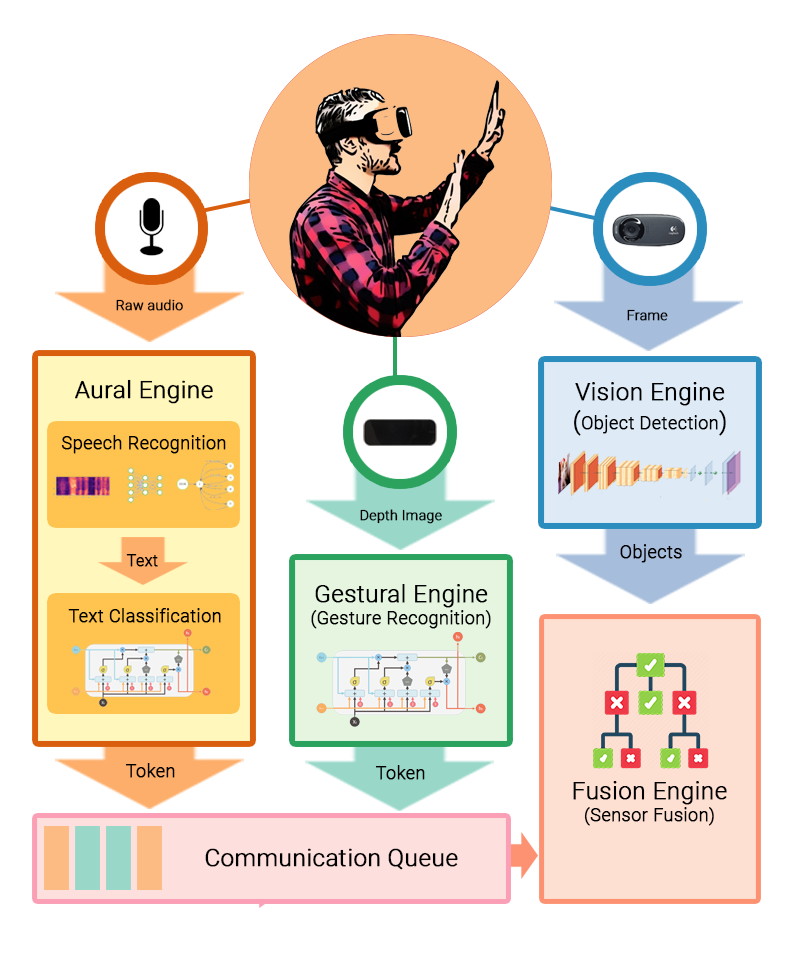}
  \caption{High-level fusion architecture}
  \vspace{-0.2 in}
  \label{fig:highlevel}
\end{figure}

Figure~\ref{fig:highlevel} illustrates the functional architecture of our proposed fusion framework. The outputs of each modality-specific Engine are exchanged via a common Communication Queue, which performs soft timestamp matching of token sequences output by each engine before feeding them into the Fusion Engine. As mentioned, state-of-the-art methods for each Engine often include complex DNN models, thus presenting significant challenges for low-latency execution on resource-constrained devices. To study the interplay between these different Engines, we shall test our framework with several different combinations of Engine-specific  models and identify the resulting tradeoff between overall system accuracy and latency. We now further describe the individual components (Engines) of our overall system.  

\textit{Vision engine:} The purpose of this subsystem is to recognize the objects with which the user interacts. This component takes raw images as input and feeds them to an object detection algorithm. Afterward, its predictions are relayed to the Fusion engine. Once the object is identified by the object detection algorithm, the objects are tracked across successive frames by an object tracker, namely KCF (Kernelized Correlation Filter)~\cite{Henriques2015}, which helps to reduce the computational overhead.

\textit{Aural engine:} This subsystem is responsible for transcribing users' utterances and classifying them to infer the user's action (or command). The speech modality is considered to be the primary input interaction modality of our system. An Automatic Speech Recognition (ASR) algorithm consumes the pre-processed captured audio signal before being fed into a text classification algorithm, which requires encoding text into a data structure appropriate for both training and inference. In order to extract the object with which user interacts, the utterance is passed through a hash table in which the keys are the names of objects and values are pre-defined identifiers. Multiple keys can have the same identifier since an object can have synonyms (e.g., mobile phone or smart phone). Afterward, a token consisting of the following three attributes is sent to Fusion engine: (1) \textit{operation-id} which is an identifier for the intended operation, (2) \textit{object-id} which is an identifier for the object, and (3) \textit{multiplicity} which denotes whether or not the user mentioned multiple objects.

\textit{Gestural engine:} Humans typically use hand gestures to supplement and disambiguate their vocal instructions. Accordingly, the gestural engine's goal is to classify gestures not only to identify interaction primitives, but to also support the refinement of issued voice commands. The vision-based devices are widely used in gesture recognition system as the contact-based methods can be detrimental to health~\cite{Kareem2012}. Therefore, we choose to implement this engine with a vision-based device. This engine currently uses a Leap Motion Controller~\footnote{\url{https://www.ultraleap.com/product/leap-motion-controller}} to capture depth images of the hand, which are then processed by a classifier to identify the gesture performed by the user. Finally, the inferred gesture label is relayed to the Fusion engine via a token.

\textit{Fusion engine:} This subsystem performs fusion of the three modalities mentioned above to identify the interaction performed. In our current framework, this fusion is triggered via the inputs of Aural and Gestural engines. Both of them will enqueue a token into a synchronized queue, with the tokens being subsequently consumed by the Fusion engine.  The Fusion engine triggers the Object detector in the Vision engine to identify the relevant objects. The Fusion engine waits for 5 secs (this is configurable) for the output of the Vision engine. The objects returned by the Vision Engine (these objects have the same taxonomy as the objects identified by the Aural Engine) are compared with the object identifiers from the Aural and Gestural engines to identify a matching target. However, if the primary object detection algorithm fails to identify the corresponding object, the Fusion engine signals the Vision engine to switch to another higher-accuracy, high-latency alternative object detector.  Once the corresponding (target) object has been determined, the Fusion engine signals the Vision engine to continue tracking this object.

\label{sec:architecture}

\section{Test Environment and Implementation}
To effectively demonstrate the capabilities of this architecture and study the tradeoffs between different instruction modalities, we decided to narrow down the implementation to fit a constrained scenario---one involving interactions in a \emph{class/study room}. By carefully studying the operations and activities that would be possible in such a well-defined context, we would be able to first select a set of suitable modality-specific classification models and then study the individual and coupled accuracy-vs.-latency tradeoffs. In the selected scenario:
\begin{itemize}[leftmargin=*]
    \item The vision-based object detection pipeline is optimized to detect generic items typical of a classroom environment, including a laptop, keyboard, monitor, mobile phone, book, bottle and cups. 
    \item The size of the vocabulary of the utterances, for the aural engine, is set to 188 (after filtering the stop words of a set of utterances that we gathered to describe and locate the aforementioned objects in different ways). 
    \item The system focuses on recognizing 4 distinct gestures: (1) pointing, (2) zoom-in, (3) zoom-out, and (4) capture.
\end{itemize}
 
Our implementation supports 4 main operations as depicted in Table~\ref{tab:fusion_operations}.

\begin{table}[h]
    \centering
    \begin{tabular}{p{1cm} p{1.3cm} p{2.7cm} p{2cm}}
        \toprule
        \textbf{Operation} & \textbf{Interaction Modalities} & \textbf{Use-case} & \textbf{Output} \\
        \midrule
        Locate &
        Visual and speech & 
        Searching for an object in the real environment with an utterance. &
        Identified object highlighted in the video feed. \\
        \midrule
        Describe &
        Visual, speech, and gestural &
        Learn more about an object with an utterance and/or a pointing gesture &
        Several properties of an object are superimposed on the display. \\
        \midrule
        Zoom &
        Visual and gestural &
        Zoom into the frame to make interactions with smaller objects easier &
        The current video feed is enlarged. \\
        \midrule
        Capture &
        Visual and gestural &
        Saving the current frame(s) for future reference &
        An instance of the video feed is saved as an image. \\
        \bottomrule
    \end{tabular}
    \vspace{0.05in}
    \caption{Fusion operations}
    \label{tab:fusion_operations}
     \vspace{-0.2in}
\end{table}

\noindent \textbf{Locate:} ``Locate'' operation is performed when the user wants to search for an object in the real world, within the visual frame of the camera.

\noindent \textbf{Zoom: } The "Zoom" operation is a fusion of visual and gestural modalities, which together result in an enlargement of the video feed displayed on the wearable device. While most past work on MR applications involve human interactions with nearby objects, it is known that interactions with distant objects is more challenging~\cite{Whitlock2018}. To overcome instructional ambiguity for such distant objects, the Zoom gesture is used to enlarge an appropriate portion of the video frame.

\noindent \textbf{Describe:} In the "Describe" operation, a user can perform an action, using speech and/or gestural modalities, to elicit a descriptive feedback from the MR application. The execution branches depending on the semantics of the speech command (i.e. whether the command is explicit or ambiguous). Ambiguous \textit{Describe} commands may come about with the use of the word \textit{this}, in which case the system will automatically look for gestural inputs. For example, the user may utter \textit{“What is this book”}, together with a corresponding pointing gesture. In this case, our system first identifies the inherent ambiguity in the referring expression ``this``, and then incorporates the sensed pointing gesture to identify the referred object. Another way in which a command can be deemed ambiguous depends on the speech command coupled with the environmental context. This entire fusion process is executed in three steps:

\begin{enumerate}
    \item \textit{Inferring speech command:} The token passed by Aural engine consists of the object identifier and its multiplicity. This operation can be implicitly ambiguous when the utterance includes \textit{this} keyword, hence there are two types of \textit{operation-ids} for the describe operations. 
    \item \textit{Localizing the object:} The vision engine scans the current video feed to find the object(s) specified by the token output of Aural Engine. The detection of the hand is necessary in cases where the user has issued an ambiguous command, thus scanning for elements of two object classes, (a) the object and (b) the hand of the user, is needed. In such cases, the nearest (relevant) object is estimated by comparing its relative distance to the hand if a pointing gesture is detected (by examining the token passed by Gestural engine).
    \item \textit{Resolving the ambiguity:} The overall command can either be  explicit operation (in which the user's intention is executed as it is) or ambiguous (in which additional information is required to perform correct comprehension). Ambiguity in the speech command may occur due to mismatch between the multiplicity of an object mentioned in an utterance and the number of instances of the particular object detected by the Visual Engine. When the user mentions only one instance of an object but the Visual Engine detects multiple object instances, disambiguation of the utterance is performed with the aid of the pointing gesture.
\end{enumerate}

\noindent \textbf{Capture:} The Capture command is a relatively trivial one, that causes the system to record and archive of the video feed for future reference.

\subsection{Implementation Details}
Our sensor fusion architecture is designed to be eventually deployed on wearable head-mounted display (HMD) smartglasses. For our current studies, which focus on exploring the tradeoff space across different modalities, we execute the developed framework on a desktop-class machine. All our experimental studies (results to be detailed in Section~\ref{sec:results}) are conducted on a desktop machine with an Intel Core i5-8300H processor (a 4-core CPU which can operate upto 4 GHz), 16 GB of RAM, and one Nvidia GTX 1060 GPU with 6GB of VRAM.

Our code is implemented in Python. The various DNN components are implemented using Tensorflow\footnote{\url{https://www.tensorflow.org}}, while the other non-DNN ML models are implemented with Scikit-learn\footnote{\url{https://scikit-learn.org/stable/}}. Additionally, we use OpenCV\footnote{\url{https://opencv.org}} to execute the image processing tasks, NLTK\footnote{\url{https://www.nltk.org}} to preprocess the inputs for text classification, and PyKaldi\footnote{\url{https://pykaldi.github.io/}} as a Python wrapper for Kaldi\footnote{\url{https://kaldi-asr.org/}} to implement on-device speech recognition. 

\label{sec:implementation}

\section{Experiments}
Our overall system is designed to be modular and configurable--i.e., allow the specific instance of each individual component  (e.g., speech recognition, object detection algorithms) to be modified, without affecting the overall fusion logic. To evaluate the interplay between the performance profiles of each component and the performance of the overall fusion engine, we adopt a ``divide-and-conquer" approach \cite{Dumas2009}, where we first evaluate each subsystem independently and then study the results of their coupled interaction. 

\subsection{Subsystem Evaluation}
To understand the tradeoff between accuracy and computational overhead, we choose two representatives for each such sub-system/component: \textbf{H} represents a model instance with higher computational requirements and likely higher accuracy, while \textbf{L} denotes the model instance with low computation requirements. 


\subsubsection{Object detection} In general, two-stage detectors outperform one-stage detectors in terms of detection accuracy, but they are computationally intensive for wearable platforms~\cite{Liu2020}. Therefore, we chose two state-of-the-art one-stage object detection algorithms, namely, SSD~\cite{Liu2016} and YOLO v3~\cite{Redmon2018}, and empirically evaluated their performance based on two parameters: \emph{speed}, defined as the number of processed frames per second (FPS), and \emph{accuracy}, defined by mean average precision (mAP) of object detection. We first employ transfer learning to re-train the detection models (which were initially trained for the \textit{COCO} dataset \cite{coco}) with labeled data matching our class/study environment. We manually prepared a nearly balanced dataset comprising 2095 instances of 10 objects (laptop, keyboard, mouse, monitor, mobile phone, bottle, cup, pen, book, and hand); this dataset allows the object detector to be trained to recognize a sufficiently diverse, but finite number of object classes.

Table~\ref{tab:objectdetection} depicts the performance of re-trained models. YOLO v3 is regarded as a high computational (\textbf{H}) model, as it has significantly higher latency, but  outperforms the other algorithm in terms of accuracy. SSD Inception V2 model has been chosen as a low computational (\textbf{L}) model as it can achieve a higher throughput (higher number of frames per second), albeit at the expense of a lower accuracy.

\begin{table}[t]
\centering
\begin{tabular}{lcc}
\toprule
\textbf{Algorithm} &
\textbf{mAP} &
\textbf{FPS} \\
\midrule
SSD Inception v2 & 0.797 & 39 \\ 
YOLO v3 & 0.963 & 12 \\
\bottomrule
\end{tabular}
\vspace{0.05in}
\caption{Results obtained from object detection algorithms}
\label{tab:objectdetection}
\vspace{-0.2in}
\end{table}

\subsubsection{Automatic Speech Recognition} The resources of a wearable device (the eventual target for our proposed MMI system) will be significantly constrained, especially for multimodal comprehension where multiple ML models will need to be executed concurrently. 
Accordingly, we explore various alternatives for ASR (automatic speech recognition) that exhibit moderate to high accuracy and relatively low computational overhead. The algorithms that were tested include Deepspeech\footnote{\url{https://github.com/mozilla/DeepSpeech}}, Kaldi\footnote{\url{https://kaldi-asr.org/}}, and IBM Watson Speech-to-text\footnote{\url{https://www.ibm.com/cloud/watson-speech-to-text}} on the cloud. 

Table~\ref{speech} tabulates our experimental results. Note that tools like IBM Watson offload the computation to the cloud and are thus suitable for resource-constrained devices, even though the additional round-trip propagation latency is likely to cause perceptible lag in an interactive system, especially under limited network bandwidth.  Accordingly, we designate IBM Watson as the \textbf{L} model with low computational overhead, while Kaldi, which provides a relatively high accuracy (low word error rate (WER)) with relatively low transcription time but imposes higher local computation, is chosen as the \textbf{H} model. The use of Kaldi along with the extended ASpIRE model\footnote{\url{http://kaldi-asr.org/models/m1}} provides good results with an acceptable latency of processing. This model was modified by creating our own dictionary and language model using context specific grammar and merging them with the original dictionary and language models of ASpIRE. The resulting enhanced model increased the likelihood of recognizing the context-specific words for our scenario. 


\begin{table}[t]
\centering
\begin{tabular}{@{}l c c@{}}
\toprule
 & \textbf{WER} & \textbf{Transcription Time}\\
\midrule
DeepSpeech & 13.59\% & 2.9s\\
Kaldi & 15.60\% & 1.5s\\
Modified Kaldi & 12.50\% & 1.5s\\
IBM Watson Speech-to-Text &  5.50\% & 2.5s\\
\bottomrule
\end{tabular}
\vspace{0.05in}
\caption{Performance of speech detection algorithms}
\label{speech}
\vspace{-0.1in}
\end{table}

\subsubsection{Text Classification}

To support high-accuracy text classification, we developed a modified Long-Short Term Memory (LSTM) neural model as LSTM-based models have proven to be extremely popular for text analysis~\cite{li2018news}. We adopted a word embedding layer which learns the multidimensional mapping of the input vector. Each test utterance is tokenized using word counts learned in the training process (the training process uses the data from the speech engine to learn a dictionary where the counts of words are stored) which are then fed into trained DNN. Confirming results reported previously, we observed that the sequential approach is superior to a simpler bag-of-words (BoW) approach. For example, for an utterance such as  \texttt{``Show me the details of the book''}, the LSTM model correctly classifies the instruction as \texttt{Describe}, whereas a BoW model would incorrectly label this as a \texttt{Locate} instruction. While our LSTM model is not perfect, we find that the accuracy is sufficiently high for our MMI scenarios, where utterances will be further \emph{disambiguated} by additional complementary modalities. The model was trained with a dataset consisting of 700+ utterances belonging to 3 distinct operations: (1) locate, (2) describe and (3) no\_op, with  no\_op representing a `null' class to capture non-command utterances (which arise as the system is continuously listening to the user). 
As an alternative to the LSTM model, we also tested two alternative modes: (a) a vanilla neural network (NN) and (b) a Support Vector Machine (SVM) model as well. Table~\ref{table:gesture_classification_models_results} tabulates the results. As the precision and recall of NN is much lower compared to the other two models, we then settle on the LSTM and SVM models as the \textbf{H} and \textbf{L} models, respectively.


\begin{table}[t]
    \centering
    \begin{tabular}{@{}l c c c@{}}
    \toprule
        \textbf{Algorithm} &
        \textbf{Precision} &
        \textbf{Recall} &
        \textbf{F1-Score} \\
    \midrule
    SVM (Linear Kernel) & 0.927 & 0.915  & 0.920 \\
    NN & 0.910 & 0.486 & 0.647 \\
    LSTM &  0.978 & 0.973 & 0.977 \\
    \bottomrule
    \end{tabular}
    \vspace{0.05in}
    \caption{Accuracies of text-classification models}
    \label{Tab:results}
    \vspace{-0.2 in}
\end{table}

\subsubsection{Gesture recognition} We collected data from 20 users performing a total of 260 samples of 4 distinct gestures: pointing, zooming in and out, and capture. Using this dataset, we formulated two features, namely \emph{single-finger} (euclidean distance between the palm center and the fingertip of each finger) and \emph{double-finger} features (euclidean distances between adjacent fingertips) using an approach similar to what is presented in~\cite{Lu2016}. Based on the observed time taken to perform a gesture (mean=487ms, std=197ms) and the sampling frequency (145 Hz) of Leap Motion Controller, we aggregate features from 30 samples and feed this aggregated feature set into an LSTM-based classifier. As before, the LSTM model is compared with two other models: (a)  a Support Vector Machine (SVM) and (b) a vanilla Neural Network (NN). Although the accuracy of LSTM and NN models (detailed in Table~\ref{table:gesture_classification_models_results}) are roughly comparable, LSTM was chosen over NN because of its higher precision. LSTM and SVM models were similarly chosen as \textbf{H} and \textbf{L} model, respectively.


\begin{table}[t]
\centering
\begin{tabular}{@{}l c c c c @{}}
\toprule
    \textbf{Algorithm} &
    \textbf{Precision} & 
    \textbf{Recall} &
    \textbf{F1-Score} &
    \textbf{Latency} \\
\midrule
    LSTM NN & 0.78 & 0.77 & 0.77 & 2.2ms \\
    NN & 0.77 & 0.77 & 0.77 & 1.5ms \\
    SVM (Linear kernel) & 0.74 & 0.72 & 0.72 & 0.1ms \\
\bottomrule
\end{tabular}
\vspace{0.05in}
\caption{Performance of gesture classification models}
\label{table:gesture_classification_models_results}
\vspace{-0.2in}
\end{table}

\begin{table*}[!tbh]
\centering
\resizebox{1\linewidth}{!}{
\begin{tabular}{cccccccccccccccc}
\toprule
\multicolumn{4}{c}{\multirow{2}{*}{\textbf{Algorithm}}} & \multicolumn{4}{c}{\textbf{Context A}}                                                 & \multicolumn{4}{c}{\textbf{Context B}}                                                 & \multicolumn{4}{c}{\multirow{2}{*}{\textbf{Resource Consumption}}} \\ \cline{5-12}
\multicolumn{4}{c}{}                                    & \multicolumn{2}{c}{\textbf{Latency (msecs)}} & \multicolumn{2}{c}{\textbf{Accuracy (\%)}} & \multicolumn{2}{c}{\textbf{Latency (msecs)}} & \multicolumn{2}{c}{\textbf{Accuracy (\%)}} & \multicolumn{4}{c}{}                                               \\ \hline
\textbf{OD}  & \textbf{ASR} & \textbf{TC} & \textbf{GR} & \textbf{O1 (Loc.)}         & \textbf{O2 (Desc.)}          & \textbf{O1 (Loc.)}          & \textbf{O2 (Desc.)}         & \textbf{O1 (Loc.)}         & \textbf{O2 (Desc.)}         & \textbf{O1 (Loc.)}          & \textbf{O2 (Desc.)}         & \textbf{RAM}   & \textbf{CPU}   & \textbf{GPU}   & \textbf{VRAM} \\
\midrule
H & H & H & H & \multirow{2}{*}{1209} & 1276 & \multirow{2}{*}{90.2} & 87.0 & \multirow{2}{*}{1255} & 1290 & \multirow{2}{*}{77.8} & 72.0  & 6217 & 38 & 32 & 5657 \\
H & H & H & L &                       & 1265 &                       & 77.3 &                       & 1282 &                       & 64.0  & 6155 & 31 & 31 & 5597 \\
H & H & L & H & \multirow{2}{*}{1152} & 1187 & \multirow{2}{*}{90.2} & 87.0 & \multirow{2}{*}{1187} & 1194 & \multirow{2}{*}{77.8} & 72.0  & 5949 & 31 & 31 & 5597 \\
H & H & L & L &                       & 1180 &                       & 77.3 &                       & 1185 &                       & 64.0  & 5890 & 31 & 31 & 5753 \\
H & L & H & H & \multirow{2}{*}{3171} & 3296 & \multirow{2}{*}{60.0} & 75.0 & \multirow{2}{*}{2864} & 3089 & \multirow{2}{*}{49.6} & 60.0  & 5195 & 35 & 23 & 5658 \\
H & L & H & L &                       & 3253 &                       & 66.7 &                       & 3140 &                       & 53.3  & 5110 & 34 & 24 & 5597 \\
H & L & L & H & \multirow{2}{*}{3102} & 3190 & \multirow{2}{*}{60.0} & 75.0 & \multirow{2}{*}{3048} & 3149 & \multirow{2}{*}{49.6} & 60.0  & 4904 & 35 & 24 & 5597 \\
H & L & L & L &                       & 3211 &                       & 66.7 &                       & 3007 &                       & 53.3  & 4854 & 34 & 19 & 5755 \\
L & H & H & H & \multirow{2}{*}{846}  & 919  & \multirow{2}{*}{71.5} & 69.0 & \multirow{2}{*}{860}  & 877  & \multirow{2}{*}{52.9} & 51.0  & 3479 & 53 & 17 & 3479 \\
L & H & H & L &                       & 916  &                       & 61.3 &                       & 859  &                       & 45.3  & 3411 & 53 & 17 & 3617 \\
L & H & L & H & \multirow{2}{*}{789}  & 844  & \multirow{2}{*}{71.5} & 69.0 & \multirow{2}{*}{782}  & 784  & \multirow{2}{*}{52.9} & 51.0  & 3197 & 53 & 18 & 3617 \\
L & H & L & L &                       & 834  &                       & 61.3 &                       & 780  &                       & 45.3. & 3170 & 51 & 17 & 3556 \\
L & L & H & H & \multirow{2}{*}{2727} & 3001 & \multirow{2}{*}{52.2} & 60.0 & \multirow{2}{*}{2656} & 2584 & \multirow{2}{*}{36.5} & 42.0  & 2444 & 37 & 15 & 3678 \\
L & L & H & L &                       & 3065 &                       & 53.3 &                       & 2665 &                       & 37.3  & 2383 & 37 & 15 & 3617 \\
L & L & L & H & \multirow{2}{*}{2774} & 3007 & \multirow{2}{*}{52.2} & 60.0 & \multirow{2}{*}{2664} & 2602 & \multirow{2}{*}{36.5} & 42.0  & 2172 & 39 & 18 & 3617 \\
L & L & L & L &                       & 2745 &                       & 53.3 &                       & 2552 &                       & 37.3  & 2116 & 39 & 18 & 3554 \\
\bottomrule

\end{tabular}}
\vspace{0.05in}
\caption{Performance comparison of different model combinations \begin{footnotesize}(OD[L] - SSD, OD[H] - YOLO, ASR[L] - IBM Watson, ASR[H] - Kaldi, TC[L] - SVM, TC[H] - LSTM, GR[L] - SVM, GR[H] - LSTM)\end{footnotesize}}
\label{tab:latency_comparison}
 \vspace{-0.2in}
\end{table*}

\subsection{Fusion-based System Evaluation}
The performance of the sensor fusion architecture can be modeled as a function of 4 algorithms, namely, object detection (OD), automatic speech recognition (ASR), text classification (TC), and gesture recognition (GR). We chose two models (\textbf{H} and \textbf{L}) for each algorithm and evaluated the latency of the system (time taken to display the output for 5 frames from the start of an action). If the system utilizes a cloud-based service, the latency figures include the additional overhead of network communication. To compute latency, two fusion operations, Locate (O\textsubscript{1}) and Describe (O\textsubscript{2}), are considered as they require multimodal interactions, whereas Zoom and Capture operations are determined solely via gesture recognition. Each operation was repeated 30 times--i.e, the study involved 30 different utterances with different objects, for both the Describe and Locate operations).

Because the comprehension accuracy is a function of not just the choice of algorithms but also the \emph{environmental context}, we evaluated the performance of different fusion strategies under two distinct contexts:
(a) \textbf{Context A} in which 5 objects (4 large objects and 1 small objects) are placed at a distance of 1 meter from the observer; and (b) \textbf{Context B} in which 5 different objects (3 large and 2 small objects) are placed at a distance of 2 meters. To understand the impact of different algorithm choices (\emph{H} vs. \emph{L}) in detail, we also measured the resource consumption in terms of usage of RAM, VRAM, CPU and GPU resources. 

Table~\ref{tab:latency_comparison} tabulates the observed latency and accuracy values, for $2^4=16$ different combinations of the individual model pipelines, as well as the corresponding consumption of computing resources. Note that the Locate primitive ($O1$) does not involve the use of a gestural input; accordingly, the latency and accuracy numbers for $O1$ are evaluated only for $2^3=8$ distinct combinations. We make the following key initial observations:
\begin{itemize}[leftmargin=*]
    \item The overall latency of the fusion process is roughly constant across different contexts and operator primitives, given a specific combination of (OD, ASR, TC, GC) models. However, the latency exhibits significant variation across different model choices--e.g., for $O1$, the latency can vary from approx. 850 msecs (for the (L,H,H) tuple) to over 2700 msecs (for the (L,L,L)) combination. Interestingly, the choice of $ASR=L$ results in a significant increase in computation latency, due to the added network latency of interacting with the cloud-based ASR engine.
    
    \item While there is still an overall trend of a tradeoff between overall complexity and accuracy, the tradeoff is not linear in the performance of individual components, but varies due to the coupling across the different modalities. For example, for the case of the $O2$(=Describe) primitive, the accuracy numbers for (H,H,L,H) and (H,H,H,H) are both approximate equal, reaching a value of 87\% for context A and 72\% for context B, respectively. The choice of a lower complexity TC model does, however, result in significantly lower resource overhead--e.g., the CPU utilization reduces by $\sim$7\%--and a nearly 100msec reduction in latency. Clearly, in this case, $TC=L$ is a preferred choice, as it results in lower processing overhead and latency without any concomitant negative impact on system accuracy. 
\end{itemize}

\begin{figure}
    \vspace{-0.05in}
    \begin{subfigure}[b]{0.48\columnwidth}
        \includegraphics[width=\linewidth]{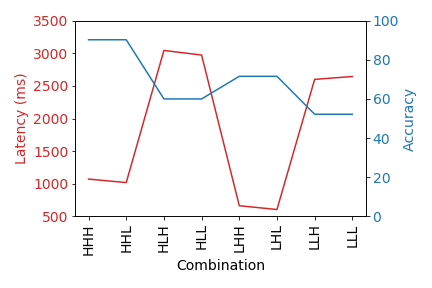}
        \caption{Context A}
    \end{subfigure}
    \hfill
    \begin{subfigure}[b]{0.48\columnwidth}
        \includegraphics[width=\linewidth]{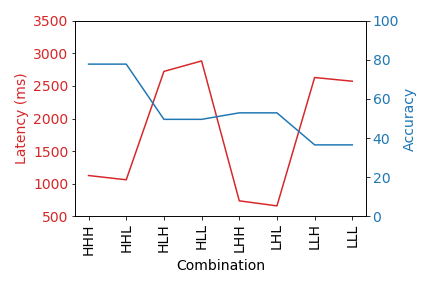}
        \caption{Context B}
    \end{subfigure}
    \caption{Tradeoff between accuracy and latency in \textbf{locate} operation}
    \label{fig:tradeoff_locate}
    \vspace{-0.2in}
\end{figure}

\begin{figure}
    \vspace{-0.05in}
    \begin{subfigure}[b]{0.48\columnwidth}
        \includegraphics[width=\linewidth]{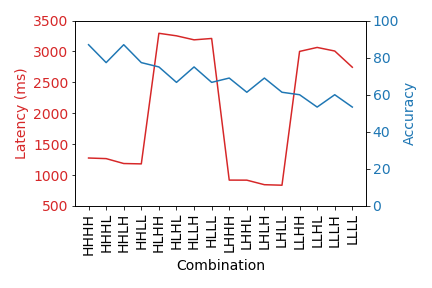}
        \caption{Context A}
    \end{subfigure}
    \hfill
    \begin{subfigure}[b]{0.48\columnwidth}
        \includegraphics[width=\linewidth]{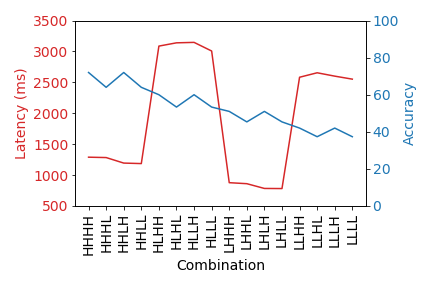}
        \caption{Context B}
    \end{subfigure}
    \caption{Tradeoff between accuracy and latency in \textbf{describe} operation}
    \label{fig:tradeoff_describe}
    \vspace{-0.2in}
\end{figure}

\textbf{Latency vs. Accuracy for different Contexts:} We next study the impact of different environmental contexts on this overall latency-vs.-accuracy tradeoff. Figures~\ref{fig:tradeoff_locate} and~\ref{fig:tradeoff_describe} plot the variations in accuracy and latency for both contexts A and B, for the cases of the Locate \& Describe operators, respectively. We can see that the relative performance is \emph{context dependent}. For example, context B (which has a more distant view of objects) observes a steeper drop in accuracy as we progressively select less complex models. In particular, with the increased distance to the observed objects (context B), the accuracy of the SSD (OD=L) algorithm  becomes significantly inferior to that obtained by the use of YOLO (OD=H). On the other hand, both $TC=L$ and $TC=H$ models show similar performance in terms of accuracy across different contexts, implying that the SVM-based classifier (TC=L) model is preferable for both contexts. The overall latency variation also differs across contexts--e.g., for the Locate operator, the (H,L,L) combination incurs higher latency for Context B but lower latency for Context A, compared to the (H,L,H) combination.

\noindent \textbf{Latency vs. Accuracy for Different Operators:} Figures~\ref{fig:tradeoff_locate} and~\ref{fig:tradeoff_describe} also help us study and evaluate the impact of different model choices, for different interaction operations. We see that the Describe operation experiences a higher latency (almost 100 msecs higher for the (H,L,H,H) combination), compared to the latency experienced by an identical model applied to the Locate operation. Moreover, there are significant differences in the resulting accuracy. In general, for an identical selection of model choices, the Describe operation has approx. 10-15\% lower accuracy. However, most interestingly, this degradation in accuracy is not universal across all model choices. Note, for example, that (H,L,L,H) has approx. 15\% higher accuracy for the Describe operation, compared to the corresponding accuracy for the Locate operator. In this case, the additional information provided by the recognition of the pointing gesture helps to isolate the target object much more precisely, without a significant increase in the overall processing latency.



\textbf{Effect of resource utilization on latency:} From the results obtained and plotted in Table~\ref{tab:latency_comparison}, it is clear that the choice of different models can result in very different profiles of resource consumption as well. For example, we observe that for the object recognition pipeline, the YOLO model (OD=H) utilize significantly more resources (49\% in RAM, 35\% in GPU, and 36\% in VRAM) compared to SSD (OD=L), but with lower CPU usage (35\%). Consequently, the choice of SSD seems to reduce the latency by 17 - 22\%. Similarly, the use of a cloud offloading for ASR results in a significantly higher processing latency (approx. 180-200\% higher), due to the network latency. However, the resource consumption in terms of RAM (24\%) and CPU (10\%) are slightly lower compared to the alternate on-device implementation. 

\noindent \textbf{Key Takeaways:} Our detailed performance results demonstrates the absence of \emph{an universally superior} combination of models--the choice of models and the consequent accuracy vs. latency tradeoff depends on the complexity of the environmental context and the interaction primitive being performed. For the specific \emph{Study Room} setting that we analyzed, 3 combinations emerge as suitable candidates under different contexts: (1) \texttt{HHLH}, a combination that imposes high CPU and GPU overhead, but is suitable for scenarios where high accuracy and low latency are essential, (2) \texttt{LHLH}, which is a CPU-intensive (high RAM and CPU usage and low GPU usage) combination and is suitable for contexts where latency is prioritized over the accuracy, and (3) \texttt{LLLH} which is especially appropriate for situations where the accuracy and latency constraints are less stringent. 





\label{sec:results}

\section{Discussion}
While our work establishes the non-obvious coupling between the computation pipelines of different modalities (and the resulting latency vs. accuracy tradeoffs), there are, however several issues that need additional exploration.

\noindent \textbf{Automatic Inference of Context (Scene Complexity):} Our work shows that the right choice of model combinations is \emph{context dependent}. However, for practical application of context-based model selection, the context itself must first be determined. Context determination itself (e.g., whether the current physical scene is cluttered or only has a few objects) requires additional computation, and the  benefit of context-dependent model adaptation may be negated if the context determination process itself has high complexity and latency. Accordingly, developing low-complexity, lightweight complexity estimators is an important prerequisite for our proposed operational model.

\noindent \textbf{Incorporating History in the Interaction Comprehension Pipeline:} Our current experimentation settings and results  utilize a \emph{memoryless} interaction model, where each interaction primitive is analyzed and estimated in isolation. In real environments, interaction primitives are obviously temporally correlated--e.g., a Zoom operation is likely to be followed by a Describe operation. The incorporation of such priors (likelihoods) is likely to further improve the process of dynamic model selection.

\noindent \textbf{Incorporating Additional Metrics in the Overall Tradeoff:} Our current evaluation has concentrated primarily on the accuracy vs. latency tradeoff. While these are important system parameters, the choice of models may need to consider additional metrics as well. For example, the energy overhead is likely to be another important constraint, especially when battery-constrained wearable devices are used continuously (e.g., in smart factories or warehouses). As our results demonstrate, latency itself may be distinct from resource consumption--e.g., the cloud-based ASR model has higher latency due to network overheads, but consumes lower CPU and RAM resources. 

\section{Conclusion}
In this paper, we focused on the problem of developing a suitable multimodal sensing and fusion framework to support natural interactivity for mixed reality (MR) applications. Our work is motivated by the rapid growth in use of DNN-based complex models for sensing-based comprehension and perception tasks, and the challenge of executing them on resource-limited pervasive devices. We demonstrate that multimodal fusion may offer a way to reduce the reliance on such complex DNN models---high accuracy estimation on one or more sensing modalities may allow the overall accuracy of fusion to remain unaffected, even if the other sensing modalities utilize less computationally-complex processing pipelines.

To support such a model, we first present a configurable and extensible multimodal sensor fusion framework, where the models for individual sensing modalities can be dynamically modified without affecting the the overall fusion process. We then study a specific instantiation of this framework, applied to a study room setting, which utilizes vision, aural and gestural input to support four different interaction primitives. By experimental studies, we quantify the performance of two different choices (low and high complexity) for each model, first in isolation and then jointly (after sensor fusion). Our studies reveal the choice of an appropriate \emph{combination} of modality-specific models is non-trivial: depending on the models chosen, the instruction comprehension latency can vary from $\sim$700-3200 msecs and the accuracy between $\sim$26-90\%. Moreover, the optimal model combination depends both on the scene complexity and the specific interaction primitive. On finer inspection, we see that (a) most optimal combinations involve the use of a higher complexity Gesture Recognizer (GR) and a low-complexity Text Classifier (TC), (b) the choice of a high vs. low complexity Object Detector (OD) is dependent on the scene complexity and (c) a higher-complexity automatic speech recognizer (ASR) is usually preferred over a lower-complexity cloud-based model, except for extremely resource-poor devices. We hope that our work motivates more careful focus on the problem of adaptive model selection for low-latency, interactive instruction comprehension.

\section*{Acknowledgements}
This research is supported by the National Research Foundation, Singapore under its International Research Centres in Singapore Funding Initiative, and Ministry of Education (MoE), Singapore under AcRF Tier-1 grant 19-C220-SMU-008. Any opinions, findings and conclusions or recommendations expressed in this material are those of the author(s) and do not reflect the views of National Research Foundation, Singapore.

\bibliographystyle{IEEEtran}
\bibliography{bibliography.bib}

\end{document}